\documentclass{JHEP3} 
\usepackage{amsmath,amssymb}
\usepackage{type1cm}
\usepackage{epsfig}
\usepackage{bbm}

\DeclareMathOperator{\tr}{tr}

\DeclareMathOperator{\Tr}{Tr}

\DeclareMathOperator{\SU}{SU}

\DeclareMathOperator{\U}{U}
\DeclareMathOperator{\OO}{O}
\DeclareMathOperator{\Real}{Re}
\DeclareMathOperator{\diag}{diag}
\newcommand{\rmd}{{\rm d}}
\newcommand{\rmD}{{\rm D}}

\def\endproof{\hskip0.6em plus0.1em minus0.1em
\setbox0=\null\ht0=5.4pt\dp0=1pt\wd0=5.3pt
\vbox{\hrule height0.8pt
\hbox{\vrule width0.8pt\box0\vrule width0.8pt}
\hrule height0.8pt}}
\newtheorem{theorem}{Theorem}[section]

\newcommand\fverb{\setbox\pippobox=\hbox\bgroup\verb}
\newcommand\fverbdo{\egroup\medskip\noindent%
                        \fbox{\unhbox\pippobox}\ }
\newcommand\fverbit{\egroup\item[\fbox{\unhbox\pippobox}]}
\newbox\pippobox

\title{Four-dimensional lattice chiral gauge theories with anomalous fermion
content}

\author{Yoshio Kikukawa\\
Institute of Physics, University of Tokyo, Komaba, Tokyo 153-8902, Japan\\
E-mail: \email{kikukawa@hep1.c.u-tokyo.ac.jp}}
\author{Hiroshi Suzuki\\
Theoretical Physics Laboratory, RIKEN, Wako 2-1, Saitama 351-0198, Japan\\
E-mail: \email{hsuzuki@riken.jp}}

\received{\today}               
\accepted{\today}               

\preprint{UT-Komaba/07-13\\RIKEN-TH-109}

\abstract{%
In continuum field theory, it has been discussed that chiral gauge theories
with Weyl fermions in anomalous gauge representations (anomalous gauge
theories) can consistently be quantized, provided that some of gauge bosons
are permitted to acquire mass. Such theories in four dimensions are inevitablly
non-renormalizable and must be regarded as a low-energy effective theory with a
finite ultraviolet (UV) cutoff. In this paper, we present a lattice framework
which enables one to study such theories in a non-perturbative level. By
introducing bare mass terms of gauge bosons that impose ``smoothness'' on the
link field, we explicitly construct a consistent fermion integration
measure in a lattice formulation based on the Ginsparg-Wilson (GW) relation.
This framework may be used to determine in a non-perturbative level an upper
bound on the UV cutoff in low-energy effective theories with anomalous fermion
content. By further introducing the St\"uckelberg or Wess-Zumino (WZ) scalar
field, this framework provides also a lattice definition of a non-linear sigma
model with the Wess-Zumino-Witten (WZW) term.}

\begin{document}

\maketitle 

\section{Introduction}
In continuum field theory, it has been discussed that chiral gauge theories
with Weyl fermions in anomalous gauge representations, the so-called
anomalous gauge theories~\cite{%
Jackiw:1984zi,Faddeev:1986pc,Krasnikov:1985bn,Harada:1986wb,Babelon:1986sv,
Niemi:1986pw,Rajaraman:1986jf,Thompson:1987cg,Miyake:1989nw,
Andrianov:1989by,Kieu:1989xi,Miyake:1991ce},
can consistently be quantized, provided that some of gauge bosons are permitted
to acquire (bare) mass~\cite{Preskill:1990fr}. Such theories in four dimensions are inevitably
non-renormalizable and must be regarded as a low-energy effective theory with a
finite UV cutoff. On the basis of perturbation theory, it has also been argued
that the UV cutoff has an upper bound given by the gauge boson mass up to a
proportionally constant~\cite{Preskill:1990fr}. See also
refs.~\cite{Sterling:1981za,D'Hoker:1984ph,Einhorn:1986za,Ball:1986wr}.

On the other hand, following a general lattice formulation of chiral gauge
theories of refs.~\cite{Luscher:1999du,Luscher:1999un}, it has been shown that
the fermion sector of a wide class of anomalous gauge theories, that includes
all four-dimensional anomalous theories, cannot consistently be defined on the
lattice~\cite{Kikukawa:2002ms,Fujiwara:2003np,Matsui:2004dc}. In appendix~A, we
present the essence of this observation in the form of a no go theorem that is
quite independent of a specific lattice formulation. (Although the theorem in
appendix~A covers only non-abelian theories, a similar no go theorem can be
established also for compact abelian theories.)
See also refs.~\cite{Neuberger:1998xn,Adams:1999um,Adams:2000yi,Adams:2001jd,%
Adams:2002ms} for related studies. Although this statement is mathematically
correct, the conclusion appears somewhat unnatural from a physical point of
view: At this moment, we know only a low-energy (compared to, say, the Plank
scale) spectrum of fermions with which the gauge anomalies are fortunately
cancelled. It is quite possible, however, that a new heavy fermion that would
give rise to an additional gauge anomaly will be discovered. Then does the
above statement imply that we have to suspend investigations of dynamics (based
on the lattice) of such a system until other heavy fermions that complete the
anomaly cancellation will be discovered? In other words, do we have to know an
anomaly-free fundamental theory very precisely to study dynamics of chiral
gauge theories at an (low) energy scale of our concern? This appears unnatural.

In this paper, we give an answer for the above question at least partially. We
show that the fermion sector of anomalous gauge theories can consistently be
formulated on the lattice, if one introduces a bare mass term of the gauge
field and imposes a sufficiently strong ``smoothness'' condition of the link
field. More definitely, we can explicitly construct a consistent Weyl fermion
integration measure in the sense of refs.~\cite{Luscher:1999du,Luscher:1999un}
in the vacuum sector of the configuration space of link fields. Technically,
such mass terms remove gauge field configurations that cause
obstructions~\cite{Fujiwara:2003np,Matsui:2004dc} for a consistent fermion
integration measure. Such restriction on gauge degrees of freedom was not
assumed in refs.~\cite{Fujiwara:2003np,Matsui:2004dc} and thus we can evade the
above conclusion on an impossibility of lattice anomalous gauge theories. See
appendix~A for a detailed account on these points.

The mass term of gauge bosons we introduce is not invariant under lattice gauge
transformations. However, the gauge invariance is anyhow broken by fermions in
an anomalous gauge representation. Introduction of a bare mass term is in fact
very natural because, as is well-known, the gauge anomaly induces mass for
gauge bosons through higher-order diagrams even if bare mass is set to be
zero~\cite{Gross:1972pv,Bouchiat:1972iq}. Since mass terms of gauge bosons
in four dimensions imply non-renormalizability, our lattice framework should be
used with finite lattice spacings. (In this paper, we consider only
four-dimensional spacetime.) In this way, at least for cases that all gauge
bosons are massive, we have a picture in lattice gauge theory that is
consistent with expectations in the continuum theory~\cite{Preskill:1990fr}.

Not only for clarifying the theoretical issue elucidated above, our lattice
framework could also be used for practical purposes. This framework may be used
to determine in a non-perturbative level an upper bound on the UV cutoff in
low-energy effective theories with anomalous fermion content. This possibility
in lattice gauge theory was first suggested in ref.~\cite{Neuberger:2000wq}. If
we further introduce the St\"uckelberg or WZ scalar field, this framework
provides also a lattice definition of a four-dimensional non-linear sigma
model with the WZW term. Thanks to a lattice Dirac operator that satisfies the
GW relation~\cite{Ginsparg:1982bj}, such as the overlap Dirac
operator~\cite{Neuberger:1998fp,Neuberger:1998wv}, the WZW term has expected
topological properties~\cite{Fujiwara:2003np}.

Throughout this paper, the spacetime dimension is set to be four. Greek
letters, $\mu$, $\nu$, \dots, run from~0 to 3. We consider a four-dimensional
square lattice
\begin{equation}
   \left\{x\in a\mathbb{Z}^4\mid 0\leq x_\mu<L\right\},
\end{equation}
where $a$~denotes the lattice spacing. A unit vector in, say, the
$\mu$-direction is denoted by~$\hat\mu$. For definiteness, the gauge group~$G$
is taken to be $\SU(N)$, but inclusion of $\U(1)$ factors and other $\SU(N')$
factors is straightforward. The standard link variables are denoted by
$U(x,\mu)\in G$. We assume that a Weyl fermion belongs to a unitary (anomalous
and generally irreducible) representation~$R$ of $G$.

\section{Lattice formulation}
\subsection{General framework}
The expectation value of an operator~$\mathcal{O}$ in our lattice framework is
defined by
\begin{equation}
   \left\langle\mathcal{O}\right\rangle
   =\frac{1}{\mathcal{Z}}\int\prod_x\prod_\mu\rmd U(x,\mu)\,
   e^{-S_{\text{G}}[U]-S_{\text{mass}}[U]}
   \left\langle\mathcal{O}\right\rangle_{\text{F}}[U],
\label{twoxone}
\end{equation}
where $\rmd U(x,\mu)$ denotes the standard Haar measure and
\begin{equation}
   \mathcal{Z}=\int\prod_x\prod_\mu\rmd U(x,\mu)\,
   e^{-S_{\text{G}}[U]-S_{\text{mass}}[U]}\,
   \left\langle1\right\rangle_{\text{F}}[U]
\label{twoxtwo}
\end{equation}
is the full partition function. The functional integration with respect to a
Weyl fermion is given by
\begin{equation}
   \left\langle\mathcal{O}\right\rangle_{\text{F}}[U]
   =\int\rmD[\psi]\rmD[\overline\psi]\,
   \mathcal{O}\,e^{-S_{\text{F}}[\psi,\overline\psi,U]}
\label{twoxthree}
\end{equation}
and, as usual, fermion fields in $\mathcal{O}$ are Wick-contracted by the
fermion propagator that can be read off from the action~$S_{\text{F}}$. What is
non-trivial is a construction of the Weyl
determinant~$\langle1\rangle_{\text{F}}[U]$ or, equivalently, a definition of the
fermion integration measure~$\rmD[\psi]\rmD[\overline\psi]$. The basic idea of
our framework is simple but a construction is somewhat complex. Thus we explain
a definition of actions part by part in following subsections. A detailed
account on the fermion integration measure (that is a crucial part of our
framework) will be given in the next section.

\subsection{Modified plaquette action $S_{\text{G}}$}
We start with a definition of the gauge action~$S_{\text{G}}$. It is given by
\begin{equation}
   S_{\text{G}}[U]=\frac{1}{g_0^2}\sum_x\sum_{\mu\nu}
   \mathcal{L}_{\mu\nu}(x),
\label{twoxfour}
\end{equation}
where $g_0$ denotes the bare gauge coupling constant and the
functions~$\mathcal{L}_{\mu\nu}(x)$ are defined by
\begin{equation}
   \mathcal{L}_{\mu\nu}(x)=
   \begin{cases}
   \frac{\displaystyle
   \Real\tr\left\{1-\mathcal{P}_{\mu\nu}(x)\right\}}
   {\displaystyle
   1-\Real\tr\left\{1-\mathcal{P}_{\mu\nu}(x)\right\}/f_R(\epsilon)}
   &\text{if $\Real\tr\left\{1-\mathcal{P}_{\mu\nu}(x)\right\}
   <f_R(\epsilon)$},\\
   +\infty&\text{otherwise},
   \end{cases}
\label{twoxfive}
\end{equation}
from the plaquette variables
\begin{equation}
   \mathcal{P}_{\mu\nu}(x)=U(x,\mu)U(x+a\hat\mu,\nu)
   U(x+a\hat\nu,\mu)^{-1}U(x,\nu)^{-1}.
\label{twowxsix}
\end{equation}
In eq.~(\ref{twoxfive}), $\epsilon$ is a constant being independent of
gauge-field configurations. The action~$S_{\text{G}}$ is a modified plaquette
action~\cite{Luscher:1999du} which dynamically imposes the restriction
$\Real\tr\{1-\mathcal{P}_{\mu\nu}(x)\}<f_R(\epsilon)$ for all $x$, $\mu$
and $\nu$ on gauge-field configurations. It can be shown that, with an
appropriate choice of the function $f_R(\epsilon)$ that depends also on a gauge
group representation~$R$ of the Weyl fermion, the restriction implies the
so-called admissibility
condition~\cite{Luscher:1981zq,Hernandez:1998et,Neuberger:1999pz}
\begin{equation}
   \left\|1-R[\mathcal{P}_{\mu\nu}(x)]\right\|<\epsilon\qquad
   \text{for all $x$, $\mu$, $\nu$}.
\label{twoxseven}
\end{equation}
In this expression, $\|A\|$ denotes the matrix norm, i.e., the square root
of the maximal eigenvalue of $A^\dagger A$, and $R$ denotes the gauge group
representation of the Weyl fermion.\footnote{For subsequent discussions, it
is useful to note the relation $\|1-U\|=\sqrt{\max_i2(1-\cos\theta_i)}$ for
a unitary matrix~$U$ with eigenvalues $e^{i\theta_i}$.} For instance, we can
take $f_R(\epsilon)=\epsilon^2/2$ for the fundamental representation of
$\SU(N)$ and $f_R(\epsilon)=\epsilon^2/8$ for the adjoint representation of
$\SU(N)$. Note that the admissibility is a gauge invariant condition. That is,
it is invariant under the gauge transformation
\begin{equation}
   U(x,\mu)\to
   U^\Lambda(x,\mu)=\Lambda(x)U(x,\mu)\Lambda(x+a\hat\mu)^{-1},
\label{twoxeight}
\end{equation}
where $\Lambda(x)\in G$.

The modified plaquette action~(\ref{twoxfour}) defines the Boltzmann
weight~$e^{-S_{\text{G}}}$ that is a product of local $C^\infty$ functions of link
variables. It differs from the standard plaquette action substantially only for
field configurations in which the field strength is of the order of the UV
cutoff, $O(1/a^2)$. Thus the modification can be regarded as a part of
allowable lattice artifacts. It is quite conceivable that the modified action
belongs to the same universality class as conventional gauge actions in the
weak coupling region.

An implication of the admissibility~(\ref{twoxseven}) is two-fold. First, it
ensures that the overlap-Dirac
operator~\cite{Neuberger:1998fp,Neuberger:1998wv}, that we will adopt below, is
well-defined and local~\cite{Hernandez:1998et,Neuberger:1999pz} if $\epsilon$
is less than $1/[6(2+\sqrt{2})]$. Second, the admissibility divides the
space of lattice gauge-field configurations into ``topological
sectors''~\cite{Luscher:1981zq}. In fact, these two facts are closely related
to each other through the lattice index
theorem~\cite{Hasenfratz:1998ft,Hasenfratz:1998jp,Hasenfratz:1998ri,%
Luscher:1998pq}. In the most part of this paper, we will consider the vacuum
sector in the space of gauge-field configurations, that is, one of topological
sectors that contains the trivial vacuum $U(x,\mu)\equiv1$. A possible
generalization to non-trivial topological sectors will briefly be mentioned at
the very end of this paper.

The space of admissible gauge fields specified by eq.~(\ref{twoxseven})
generally possesses a non-trivial topological structure. At this moment, a
parametrization of the space is known only for $G=\U(1)$~\cite{Luscher:1999du}.
This fact is one of main obstacles for a generalization of a construction of
abelian lattice chiral gauge
theories~\cite{Luscher:1999du,Kadoh:2003ii,Kadoh:2004uu,Kadoh:2005fa} to
non-abelian theories. In the present context, we can avoid this difficulty by
further restricting the space of gauge-field configurations within a ball
enclosing the trivial vacuum~$U(x,\mu)\equiv1$. This is an important role of
the mass term of gauge fields that we will explain next.

\subsection{Mass term $S_{\text{mass}}$}
We introduce a mass term of gauge fields of the form
\begin{equation}
   S_\text{mass}[U]=\frac{2m_0^2a^2}{g_0^2}
   \sum_x\sum_\mu\mathcal{M}_\mu(x),
\label{twoxnine}
\end{equation}
where
\begin{equation}
   \mathcal{M}_\mu(x)=
   \begin{cases}
   \frac{\displaystyle
   \Real\tr\left\{1-U(x,\mu)\right\}}
   {\displaystyle
   1-\Real\tr\left\{1-U(x,\mu)\right\}/f_R(\delta)}
   &\text{if $\Real\tr\left\{1-U(x,\mu)\right\}<f_R(\delta)$},\\
   +\infty&\text{otherwise}.
   \end{cases}
\label{twoxten}
\end{equation}
In eq.~(\ref{twoxten}), $\delta$ is a constant being independent of
gauge-field configurations. As the modified plaquette action~(\ref{twoxfour}),
this mass term~$S_{\text{mass}}$ dynamically imposes the
condition~$\Real\tr\{1-U(x,\mu)\}<f_R(\delta)$ and, as before, this implies the
condition
\begin{equation}
   \left\|1-R[U(x,\mu)]\right\|<\delta\qquad\text{for all $x$ and $\mu$},
\label{twoxeleven}
\end{equation}
which we will refer to as the ``smooth'' condition. Note that this condition is
\emph{not\/} gauge invariant, because it is a condition on link variables that
transform as eq.~(\ref{twoxeight}). This is not so surprising, because mass
terms of gauge bosons are anyhow not gauge invariant. The associated Boltzmann
weight~$e^{-S_{\text{mass}}}$ is a product of local $C^\infty$-class functions of
link variables. We choose the constant~$\delta$ such that
\begin{equation}
   \delta\leq
   \sqrt{2}\sqrt{1-\cos\left\{\pi/(N-1)\right\}}\qquad\text{for $G=\SU(N)$},
\label{twoxtwelve}
\end{equation}
\emph{and}
\begin{equation}
   4\delta+6\delta^2+4\delta^3+\delta^4\leq\epsilon.
\label{twoxthirteen}
\end{equation}

With the mass term~(\ref{twoxnine}), the above system is non-renormalizable in
the weak coupling expansion around the trivial vacuum $U(x,\mu)\equiv1$.
Although this action differs from the would-be standard mass term
$\Real\tr\{1-U(x,\mu)\}$ by the
factor~$[1-\Real\tr\{1-U(x,\mu)\}/f_R(\delta)]^{-1}$, the modification becomes
effective only when the gauge potential becomes the cutoff order~$O(1/a)$. Thus
it can be regarded as a part of allowable lattice artifacts.

Now, let us explain the meaning of the inequality~(\ref{twoxtwelve}). Under
this, the space of smooth gauge-field configurations, specified by
eq.~(\ref{twoxeleven}), is contractible. The space is thus topologically
trivial and looks like a ball with the ``radius'' $\delta$. This fact can be
shown by defining a one-parameter family of gauge-field configurations
\begin{equation}
   U_t(x,\mu)=[U(x,\mu)]^t\in G,\qquad0\leq t\leq1.
\label{twoxfourteen}
\end{equation}
The power $[U]^t$ of a unitary matrix~$U\in G$ can be defined as follows.
Suppose that a unitary matrix $U\in\SU(N)$ satisfies~$\|1-U\|<\delta$ and the
constant~$\delta$ fulfills the inequality~(\ref{twoxtwelve}). Then, such a
matrix can be represented as
\begin{equation}
   U=V\diag(e^{i\theta_1},e^{i\theta_2},\ldots,e^{i\theta_N})\,V^{-1},\qquad
   V\in G,
\end{equation}
where all angles~$\theta_i$ are in the \emph{open\/} interval
$-\pi<\theta_i<+\pi$. The power~$[U]^t$ with $0\leq t\leq1$ can then be defined
by
\begin{equation}
   [U]^t=V\,\diag(e^{it\theta_1},e^{it\theta_2},\ldots,e^{it\theta_N})\,V^{-1}\in G.
\end{equation}
Thus under the smooth condition~(\ref{twoxeleven}) with $\delta$ that fulfills
the inequality~(\ref{twoxtwelve}), we can unambiguously define the one-parameter
family~(\ref{twoxfourteen}). The one-parameter family~(\ref{twoxfourteen})
continuously connects any smooth gauge-field configuration~$U$ to the
trivial configuration, $U\equiv1$. Thus the space of smooth gauge field
configurations is contractible.

The meaning of the another inequality~(\ref{twoxthirteen}) is as follows. Under
eq.~(\ref{twoxthirteen}), any configuration which satisfies the smooth
condition~(\ref{twoxeleven}) is admissible, that is, it
satisfies~eq.~(\ref{twoxseven}). This can be seen by applying the Schwartz
inequality for the matrix norm to~eq.~(\ref{twoxseven}). Moreover, from the
above construction, we see that if $U(x,\mu)$ is admissible then the
one-parameter family~(\ref{twoxfourteen}) is also admissible.

These explain the origin of inequalities~(\ref{twoxtwelve})
and~(\ref{twoxthirteen}). Under eq.~(\ref{twoxtwelve}), the space specified by
the condition~(\ref{twoxeleven}) is a contractible ball. If $\delta$ fulfills
the inequality~(\ref{twoxthirteen}), the ball is moreover contained in the
space of admissible gauge-field configurations. In this way, we restrict
possible gauge-field configurations into a topologically trivial space. One
would be afraid of that such restriction is too strong, i.e., the
condition~(\ref{twoxeleven}) excises also gauge-field configurations which
become physically important in the continuum limit. If the link variables can
be expanded by the gauge potentials as
$U(x,\mu)\simeq1+aA_\mu(x)$ in the continuum limit, as we are assuming, the
condition becomes $|A_\mu(x)|\lesssim\delta/a$ in the continuum
limit. It is then clear that nothing important is lost in the $a\to0$ limit.

Recall that the restriction~(\ref{twoxeleven}) is not invariant under lattice
gauge transformations. We need such a non gauge invariant restriction to avoid
the no go theorem in appendix~A. This restriction moreover allows one to
construct a consistent fermion integration measure, while evading difficulty of
finding a precise parametrization of the space of admissible configurations.

On the other hand, for the above simple trick to work, it is clear that we have
to introduce mass terms for \emph{all\/} gauge bosons. In the present
framework, one cannot keep some of gauge bosons, those associated with an
(anomaly-free) unbroken subgroup~$H$, massless. This limits a range of
applicability of the present lattice framework.

\subsection{Fermion action $S_{\text{F}}$}
We now turn to the fermion action defined by
\begin{equation}
   S_{\text{F}}[\psi,\overline\psi,U]=a^4\sum_x\overline\psi(x)D\psi(x).
\label{twoxseventeen}
\end{equation}
The lattice Dirac operator~$D$ is assumed to satisfy the GW
relation~\cite{Ginsparg:1982bj}
\begin{equation}
   \gamma_5D+D\gamma_5=aD\gamma_5D,
\label{twoxeighteen}
\end{equation}
that implies an exact chiral symmetry on the lattice~\cite{Luscher:1998pq}.
For definiteness, we assume use of the overlap-Dirac
operator~\cite{Neuberger:1998fp,Neuberger:1998wv} in what follows.

We first introduce the modified chirality
matrix~\cite{Narayanan:1998uu,Niedermayer:1998bi}
$\hat\gamma_5=\gamma_5(1-aD)$. This operator satisfies
\begin{equation}
   (\hat\gamma_5)^\dagger=\hat\gamma_5,\qquad
   (\hat\gamma_5)^2=1,\qquad
   D\hat\gamma_5=-\gamma_5D,
\end{equation}
where the last two relations follow from the GW relation~(\ref{twoxeighteen}).
One then defines projection operators
\begin{equation}
   \hat P_\pm=\frac{1}{2}(1\pm\hat\gamma_5),\qquad
   P_\pm=\frac{1}{2}(1\pm\gamma_5).
\end{equation}
Note that the hatted projection operators~$\hat P_\pm$ depend on a gauge-field
configuration through the Dirac operator~$D$. It is then assumed that the
fermion variables are subject to the following constraints specifying the
(left-handed) chirality
\begin{equation}
   \hat P_-\psi(x)=\psi(x),\qquad\overline\psi(x)P_+=\overline\psi(x).
\label{twoxtwentyone}
\end{equation}
This construction of a lattice action of a Weyl fermion is equivalent to the
domain-wall formulation~\cite{Kaplan:1992bt} and to the overlap
formulation~\cite{Narayanan:1993wx,Narayanan:1994sk,Narayanan:1993ss,%
Narayanan:1995gw,Randjbar-Daemi:1995sq,Randjbar-Daemi:1995cq,%
Randjbar-Daemi:1996mj,Randjbar-Daemi:1997iz} of lattice Weyl fermions.

\section{Construction of the fermion integration measure}
To define the integration measure $\rmD[\psi]\rmD[\overline\psi]$ for a Weyl
fermion in eq.~(\ref{twoxthree}), one introduces orthonormal bases in the
constrained spaces~(\ref{twoxtwentyone}):
\begin{align}
   &\hat P_-v_j(x)=v_j(x),\qquad(v_k,v_j)=\delta_{kj},
\label{threexone}\\
   &\overline v_k(x)P_+=\overline v_k(x),\qquad
   (\overline v_j^\dagger,\overline v_k^\dagger)=\delta_{kj}
\label{threextwo}
\end{align}
and expand field variables as
\begin{equation}
   \psi(x)=\sum_jv_j(x)c_j,\qquad
   \overline\psi(x)=\sum_k\overline c_k\overline v_k(x).
\end{equation}
The fermion integration measure is then defined by
\begin{equation}
   \rmD[\psi]\rmD[\overline\psi]=\prod_j\rmd c_j\prod_k\rmd\overline c_k
\end{equation}
in terms of the Grassmann expansion coefficients.

In eq.~(\ref{threexone}), basis vectors~$\{v_j\}$ depend on the gauge-field
configuration through the projection operator~$\hat P_-$. However,
eq.~(\ref{threexone}) does not fix basis vectors uniquely (any unitary
transformation of $\{v_j\}$ with respect to the index $j$ leaves the constraint
invariant). This arbitrariness of basis vectors results in phase ambiguity of
the fermion integration measure that may depends on the gauge-field
configuration.\footnote{The basis vectors $\{\overline v_k\}$ in
eq.~(\ref{threextwo}) can be taken to be independent of a gauge-field
configuration and there is no ambiguity associated with their choice.} One has
to fix this ambiguity so that the locality and smoothness\footnote{Here, the
smoothness means that any expectation value in the fermion
sector~(\ref{twoxthree}) is a single-valued $C^\infty$-class function of link
variables. For anomaly-free chiral gauge theories, the phase moreover must be
consistent with the gauge invariance.}
hold~\cite{Luscher:1999du,Luscher:1999un}.

To study this problem, it is convenient to introduce the
\emph{measure term\/}~\cite{Luscher:1999du,Luscher:1999un}
\begin{equation}
   \mathfrak{L}_\eta=a^4\sum_x\eta_\mu^a(x)j_\mu^a(x),
\end{equation}
where $\eta^a(x)$ denotes a variation vector of link variables
\begin{equation}
   \delta_\eta U(x,\mu)=a\eta_\mu(x)U(x,\mu),\qquad
   \eta_\mu(x)=\eta_\mu^a(x)T^a,
\end{equation}
and $T^a$ are anti-hermitian generators of~$G$. The current~$j_\mu^a(x)$, that
is a function of a gauge-field configuration, is referred to as the measure
current.

It can be shown~\cite{Luscher:1999du,Luscher:1999un} that if a given measure
term satisfies several prerequisites, one can reconstruct the fermion
integration measure that is consistent with the locality and smoothness.
Generally, those prerequisites are quite non-trivial to be fulfilled if
the space of gauge-field configurations possesses a non-trivial topological
structure (such as non-contractible loops). In our present system, gauge-field
configurations are restricted by the smooth condition~(\ref{twoxeleven}) and
the space of allowed gauge-field configurations is topologically trivial. For
such a topologically trivial configuration space, the above prerequisites are
reduced to the locality and the local integrability. The locality here means
that the measure current~$j_\mu^a(x)$ is a local expression of link variables.
The local integrability is
\begin{equation}
   \delta_\eta\mathfrak{L}_\zeta-\delta_\zeta\mathfrak{L}_\eta
   +a\mathfrak{L}_{[\eta,\zeta]}
   =i\Tr\left\{\hat P_-[\delta_\eta\hat P_-,\delta_\zeta\hat P_-]\right\},
\label{threexseven}
\end{equation}
where variations $\eta$ and~$\zeta$ are assumed to be independent of link
variables.

In our present system, in fact, it is easy to construct a measure term that
fulfills the locality and the local integrability~(\ref{threexseven}). Denoting
the projection operator associated with the one-parameter
family~(\ref{twoxfourteen}) as
\begin{equation}
   P_t=\left.\hat P_-\right|_{U=U_t},
\end{equation}
we can adopt the following measure term
\begin{equation}
   \mathfrak{L}_\eta=i\int_0^1\rmd t\,
   \Tr\left\{P_t\left[\partial_t P_t,\delta_\eta P_t\right]\right\}.
\label{threexnine}
\end{equation}
It is clear that, from the locality of the overlap Dirac
operator~\cite{Hernandez:1998et,Neuberger:1999pz}, the measure current
associated with the above measure term is a local expression of link variables.
Thus the locality is ensured. One can also confirm that this measure term
satisfies the local integrability~(\ref{threexseven}). These allows us to
construct basis vectors~$\{v_j\}$ that are consistent with the locality and
smoothness~\cite{Luscher:1999du,Luscher:1999un}.\footnote{%
The construction proceeds as follows: One introduces a unitary
operators~$Q_t$ by the differential equation,
\begin{equation}
   \partial_tQ_t=[\partial_tP_t,P_t]Q_t,\qquad Q_0=1.
\end{equation}
One also computes the Wilson line associated with the measure term by
\begin{equation}
   W=\exp\left\{i\int_0^1\rmd t\,\mathfrak{L}_\eta\right\},\qquad
   a\eta_\mu(x)=\partial_tU_t(x,\mu)U_t(x,\mu)^{-1}.
\end{equation}
For our measure term~(\ref{threexnine}), we have $\mathfrak{L}_\eta=0$ for
$a\eta_\mu=\partial_tU_tU_t^{-1}$ and thus $W=1$. Finally, one makes a certain
choice of basis vectors $\{w_j\}$ for the vacuum $U\equiv1$. From these, basis
vectors for $U$ is given by
\begin{equation}
   v_j=Q_1w_j
\end{equation}
and the Weyl determinant~\cite{Luscher:1999un}
\begin{equation}
   \left\langle1\right\rangle_{\text{F}}[U]
   \left\langle1\right\rangle_{\text{F}}[1]^*
   =\det\left\{1-P_++P_+DQ_1D_0^\dagger\right\},
\end{equation}
where $D_0$ is the Dirac operator for $U\equiv1$.
}

In particular, a variation of the fermion effective action is given by
\begin{equation}
   \delta_\eta\ln\left\langle1\right\rangle_{\text{F}}[U]
   =\Tr\left\{\delta_\eta D\hat P_-D^{-1}P_+\right\}-i\mathfrak{L}_\eta
\end{equation}
and its integration along the path~(\ref{twoxfourteen}) gives the effective
action (see also ref.~\cite{Suzuki:1999qw})
\begin{align}
   \ln\left\langle1\right\rangle_{\text{F}}[U]
   -\ln\left\langle1\right\rangle_{\text{F}}[1]
   &=\int_0^1\rmd t\,\left.\left(
   \Tr\left\{\partial_tD\hat P_-D^{-1}P_+\right\}
   -i\mathfrak{L}_\eta\right)
   \right|_{U=U_t}
\nonumber\\
   &=\int_0^1\rmd t\,
   \Tr\left.\left\{\partial_tD\hat P_-D^{-1}P_+\right\}\right|_{U=U_t},
\end{align}
where the variation is given by
$a\eta_\mu(x)=\partial_tU_t(x,\mu)U_t(x,\mu)^{-1}$ and we have noted
$\mathfrak{L}_\eta=0$ for this variation~$\eta_\mu$. This completes a
construction of the fermion sector.

Note that our construction works even for a single four-dimensional Weyl
fermion in the fundamental representation of $\SU(2)$, that suffers from the
$\SU(2)$ anomaly~\cite{Witten:fp,Elitzur:1984kr}. It has been
shown~\cite{Bar:2000qa,Bar:2002sa} that the lattice formulation in
refs.~\cite{Luscher:1999du,Luscher:1999un} neatly reproduces the $\SU(2)$
anomaly, as a non-integrability along a non-contractible loop in the space of
admissible gauge-field configurations. As we emphasized, such a topologically
non-trivial structure is removed from our space of allowed gauge-field
configurations and this is the reason why we can construct a consistent
fermion integration measure for a Weyl fermion in any gauge representation. One
may wonder, then, whether our construction is potentially inconsistent if it is
applicable even to a single $\SU(2)$ Weyl fermion.

The key is again the presence of the mass term~(\ref{twoxnine}). It is not
invariant under any non-trivial gauge transformation. As pointed out in
ref.~\cite{Preskill:1990fr}, if the action is not gauge invariant, the
argument~\cite{Witten:fp} that shows inconsistency of a gauge theory
containing a single $\SU(2)$ Weyl fermion does not apply. One may have
well-defined expectation values if the action contains, say, mass terms of
gauge bosons. In this way, we again have a consistent picture.

\section{Introducing the St\"uckelberg or WZ scalar}
For some purposes, it is useful to introduce the degrees of freedom of
a $G$-valued St\"uckelberg or WZ scalar~$g(x)$ into our system. We multiply
eq.~(\ref{twoxone}) by unity, $1=\int\prod_x\rmd g(x)$, where
$\rmd g(x)$ is the Haar measure, to yield
\begin{equation}
   \left\langle\mathcal{O}\right\rangle
   =\frac{1}{\mathcal{Z}}\int\prod_x\rmd g(x)
   \int\prod_x\prod_\mu\rmd U(x,\mu)\,e^{-S_{\text{G}}[U]-S_{\text{mass}}[U^g]}
   \left\langle\mathcal{O}\right\rangle_{\text{F}}[U^g],
\label{fourxone}
\end{equation}
where we have made change of variables from $U(x,\mu)$ to
\begin{equation}
   U^g(x,\mu)=g(x)U(x,\mu)g(x+a\hat\mu)^{-1},
\label{fourxtwo}
\end{equation}
and used the gauge invariance of the gauge action~$S_{\text{G}}$ and of the
measure.

In this new picture, the integrand of the functional integration is gauge
invariant. That is,
$e^{-S_{\text{G}}[U]-S_{\text{mass}}[U^g]}\langle\mathcal{O}\rangle_{\text{F}}[U^g]$
\emph{is\/} invariant under the gauge transformations
\begin{equation}
   U(x,\mu)\to\Lambda(x)U(x,\mu)\Lambda(x+a\hat\mu)^{-1},\qquad
   g(x)\to g(x)\Lambda(x)^{-1},
\label{fourxthree}
\end{equation}
because the combination~$U^g$ is trivially invariant under these
transformations. This realization of gauge invariance may be regarded as
anomaly cancellation between a Weyl fermion and the WZ scalar~$g$. We may
define the WZW term~\cite{Wess:yu,Witten:tw} in lattice gauge theory by
\begin{equation}
   e^{-i{\mit\Gamma}_{\text{WZW}}[g^{-1},U]}
   =\frac{\displaystyle\left\langle1\right\rangle_{\text{F}}[U^g]}
   {\displaystyle\left\langle1\right\rangle_{\text{F}}[U]},
\label{fourxfour}
\end{equation}
where $g(x)$ is a $G$-valued scalar field (the WZ scalar).\footnote{
Our choice of the measure term~(\ref{threexnine}) and, as a result, a
definition of the lattice WZW term~(\ref{fourxfour}) are completely
identical to those of ref.~\cite{Fujiwara:2003np}. We can thus repeat arguments
of~ref.~\cite{Fujiwara:2003np} for the lattice WZW term~(\ref{fourxfour}).
In particular, we can see that the WZW term is a \emph{local\/} functional of
$g$ and~$U$ and possesses topological properties common to the
continuum~\cite{Witten:tw} even with finite lattice spacings. It can be shown
that it also has a correct classical continuum limit.} Then the Weyl
determinant~$\langle1\rangle_{\text{F}}[U^g]$ in eq.~(\ref{fourxone}) can be
expressed as $e^{-i{\mit\Gamma}_{\text{WZW}}[g^{-1},U]}\langle1\rangle_{\text{F}}[U]$.
The Weyl determinant~$\langle1\rangle_{\text{F}}[U]$ is not gauge invariant, but
a gauge variation of the WZW term compensates this breaking of the gauge
symmetry. This anomaly cancellation can also be regarded as a simplest variant
of the Green-Schwarz anomaly cancellation
mechanism~\cite{Green:1984sg,Witten:1984dg,Dine:1987xk}.\footnote{Note that,
however, in general Green-Schwarz mechanism in which an anti-symmetric tensor
field~$B$ (instead of a scalar field) cancels the anomaly, it is impossible to
take a ``unitary gauge'' that completely eliminates~$B$, because $B$ is
transformed into a Chern-Simons form by the gauge transformation, instead into
a function.}

As emphasized in ref.~\cite{Preskill:1990fr}, however, the above gauge
invariance in the new picture has no immediate consequence. In fact, as
eq.~(\ref{fourxone}) shows, the expectation value of any operator (such as
Wilson lines) is the same as that in the old picture~(\ref{twoxone}). If one
wishes, the unitary gauge can be taken in which
\begin{equation}
   g(x)\equiv1
\label{fourxfive}
\end{equation}
by using the gauge invariance in the new picture. Then the system reduces to
the old one. Thus, which picture (non gauge invariant or gauge invariant) we
take is just a matter of description.

Incidentally, on the lattice, we can always take the unitary
gauge~(\ref{fourxfive}) and this justifies in a non-perturbative level the
treatment in ref.~\cite{Preskill:1990fr} that assumes the absence of global
obstructions to set $g(x)\equiv1$.

In the new picture~(\ref{fourxone}), the mass term becomes a gauge invariant
kinetic term of the WZ scalar
\begin{equation}
   S_{\text{mass}}[U^g]
   =K\sum_x\sum_\mu\mathcal{M}_\mu(x),\qquad
   K\equiv\frac{2m_0^2}{g_0^2}a^2
\label{fourxsix}
\end{equation}
where
\begin{equation}
   \mathcal{M}_\mu(x)=
   \begin{cases}
   \frac{\displaystyle
   \Real\tr\left\{1-g(x)U(x,\mu)g(x+a\hat\mu)^{-1}\right\}}
   {\displaystyle
   1-\Real\tr\left\{1-g(x)U(x,\mu)g(x+a\hat\mu)^{-1}\right\}/f_R(\delta)}
   &\\
   &\hskip-16em
   \text{if $\Real\tr\left\{1-g(x)U(x,\mu)g(x+a\hat\mu)^{-1}\right\}
   <f_R(\delta)$},\\
   +\infty&\text{otherwise}.
   \end{cases}
\label{fourxseven}
\end{equation}
Since the Weyl determinant produces the WZW term, bosonic sector of our system
is just the gauged non-linear sigma model with the WZW
term.\footnote{See ref.~\cite{Chandrasekharan:2003wy} for a study of the
non-linear sigma model on the lattice.} Here we assume that the parameter~$K$
is sufficiently large so that the system is in the Higgs phase and the
expansion of link variables around $U\equiv1$ is justified in the weak coupling
limit, $g_0\to0$. There is a possibility that the sigma model is always in the
Higgs phase for all values of~$K$ when $\delta$ is sufficiently small. This
question is highly dynamical, especially with the presence of Weyl fermions,
and is beyond the scope of this paper.

\section{Topological sectors}
In this final section, we make a brief comment on a possible generalization of
our construction to topologically non-trivial sectors. As noted, the space of
gauge-field configurations is divided into topological sectors under the
admissibility condition~(\ref{twoxseven}). The fermion integration measure has
to be defined sector by sector~\cite{Luscher:1999du,Luscher:1999un}. A natural
generalization of the mass term is
\begin{equation}
   S_\text{mass}[U]=\frac{2m_0^2a^2}{g_0^2}
   \sum_x\sum_\mu\mathcal{M}_\mu(x),
\end{equation}
where
\begin{equation}
   \mathcal{M}_\mu(x)=
   \begin{cases}
   \frac{\displaystyle
   \Real\tr\left\{1-U_0(x,\mu)^{-1}U(x,\mu)\right\}}
   {\displaystyle
   1-\Real\tr\left\{1-U_0(x,\mu)^{-1}U(x,\mu)\right\}/f_R(\delta)}
   & \\
   &\hskip-13.5em
   \text{if $\Real\tr\left\{1-U_0(x,\mu)^{-1}U(x,\mu)\right\}<f_R(\delta)$},\\
   +\infty&\text{otherwise}.
   \end{cases}
\end{equation}
The idea is that, with this mass term, gauge-field configurations are
restricted within a ball with a radius~$\delta$ encircling a reference field
$U_0(x,\mu)$. The reference field may be a non-trivial one such as the (lattice
transcription of) instanton configuration. With $\delta$ that fulfills
eq.~(\ref{twoxtwelve}), we can introduce the one-parameter family
\begin{equation}
   U_t(x,\mu)=U_0(x,\mu)[U_0(x,\mu)^{-1}U(x,\mu)]^t\in G,\qquad0\leq t\leq1,
\end{equation}
which interpolates between the reference configuration~$U_0$ and the
configuration under consideration, $U$. In contrast to the case of the vacuum
sector, however, the inequality~(\ref{twoxthirteen}) is not enough to ensure
that the ball is contained in the space of admissible configurations specified
by eq.~(\ref{twoxseven}). The corresponding inequality must refer to~$U_0$.
Also, there seems no fundamental criterion to choose a particular reference
configuration~$U_0$ within a topological sector. From these reasons, we do not
pursuit this generalization any further in this paper.

\section{Conclusion}
In this paper, we have presented a lattice framework with which one can study
anomalous gauge theories with a Weyl fermion in an anomalous gauge
representation. By introducing mass terms for all gauge bosons that impose
smoothness on gauge degrees of freedom, we constructed a consistent fermion
integration measure in the formulation of
refs.~\cite{Luscher:1999du,Luscher:1999un} for the vacuum sector of the
configuration space of gauge fields. We argued that introduction of such (bare)
mass terms is physically natural. Also, in view of the no go theorem in
appendix~A, we have to place a certain non gauge invariant restriction on
lattice gauge-field configurations. An interesting question one can study with
the present lattice framework is an upper bound on the UV cutoff in low-energy
effective theories with anomalous fermion content that is suggested from a
perturbative analysis~\cite{Preskill:1990fr}.

In this paper, we have considered four-dimensional anomalous gauge theories.
The present framework, when applied to two dimensions, can provide a lattice
definition of the (off-critical) WZW model in two
dimensions~\cite{Witten:1983ar}. We hope to study this prospect in the near
future.

\acknowledgments
The work of Y.K.\ is supported in part by Grant-in-Aid for Scientific Research,
17540249.
The work of H.S.\ is supported in part by Grant-in-Aid for Scientific Research,
18540305, and by JSPS and French Ministry of Foreign Affairs under the
Japan-France Integrated Action Program (SAKURA).

\appendix

\section{A no go theorem for a Weyl fermion in an anomalous
representation}
The following no go theorem states that, under some assumptions, it is
impossible to construct a physically sensible lattice formulation of a Weyl
fermion for a wide class of anomalous gauge theories (that includes all
four-dimensional non-abelian theories). The spacetime dimension is set to
be~$2n$.
\begin{theorem}
Suppose that the compact gauge group~$G$ is semi-simple and $\pi_1(G)=0$ and
$\pi_{2n+1}(G)=\mathbb{Z}$. Then if a Weyl fermion belongs to an anomalous
representation~$R$ for which the leading anomaly coefficient~$A_{n+1}(A)$,
defined by
\begin{align}
   \tr\{R(F)^{n+1}\}=A_{n+1}(R)\tr\{F^{n+1}\}+(\text{factorized traces}),
\label{axone}
\end{align}
where $F$ is the field strength two-form in the fundamental representation, is
non-vanishing $A_{n+1}(R)\neq0$, the following four requirements are
incompatible to each other:
\begin{enumerate}
\item The lattice Weyl determinant reproduces the gauge anomaly in the
classical continuum limit.
\item Only gauge invariant restrictions are placed on link variables.
\item The modulus of the lattice Weyl determinant is gauge invariant.
\item The lattice Weyl determinant is a (at least) $C^2$-class function of link
variables.
\end{enumerate}
\end{theorem}
(Proof) We introduce a one-parameter family of lattice gauge
transformations~$\Lambda_t(x)\in G$ ($0\leq t\leq1$) such that
$\Lambda_0(x)=\Lambda_1(x)=1$. This one-parameter family~$\Lambda_t$ is a loop
in the space of lattice gauge transformations~$\mathfrak{G}$. From this, we
define a one-parameter family of pure-gauge link variables
\begin{equation}
   U_t(x,\mu)=\Lambda_t(x)\Lambda_t(x+a\hat\mu)^{-1}.
\label{axtwo}
\end{equation}
We then write the lattice Weyl determinant as
\begin{equation}
  \left\langle1\right\rangle_{\text{F}}[U]=r[U]e^{i\vartheta[U]}
\label{axthree}
\end{equation}
and define the winding number of the complex phase around the loop
\begin{equation}
   w=\frac{1}{2\pi}\int_0^1\rmd t\,\frac{\partial\vartheta[U_t]}{\partial t}.
\label{axfour}
\end{equation}
Because of prerequisites in the theorem and the assumption~1, we can repeat an
argument in the continuum~\cite{Alvarez-Gaume:1983cs}. It follows that (if the
lattice is fine enough) we can choose the one-parameter family~$\Lambda_t(x)$
such that the winding number~$w$ is non-trivial, $w\neq0$.

On the other hand, the space of lattice gauge transformations~$\mathfrak{G}$ is
topologically trivial and any loop $\Lambda_t$ is contractible
($\pi_1(\mathfrak{G})=0$). Therefore, there exists a two-parameter family of
lattice gauge transformations $\Lambda_{t,s}(x)$ ($0\leq s\leq1$) such that
\begin{equation}
   \Lambda_{t,s=0}(x)=1,\qquad\Lambda_{t,s=1}(x)=\Lambda_t(x).
\label{axfive}
\end{equation}
The corresponding two-parameter family of link variables
\begin{equation}
   U_{t,s}(x,\mu)=\Lambda_{t,s}(x)\Lambda_{t,s}(x+a\hat\mu)^{-1}
\label{axsix}
\end{equation}
can then be regarded as a two-disk~$D$ in the space of gauge fields.
Configurations belonging to this two-disk~$D$ are allowed configurations in a
lattice formulation under consideration, because of the assumption~2. Note also
that the one-parameter family~(\ref{axtwo}) is the boundary of the disk~$D$,
$\partial D$.

Now, from the assumptions~3, the modulus~$r$ is constant over the two-disk~$D$
(eq.~(\ref{axsix})). Combined with the assumption~4, this fact implies that
the one-form
\begin{equation}
   a(t,s)=\rmd\vartheta(t,s)
\end{equation}
is a (at least) $C^1$-class function on~$D$. However, then, from the Stokes
theorem
\begin{equation}
   w=\frac{1}{2\pi}\oint_{\partial D}a
   =\frac{1}{2\pi}\int_D\rmd a=0
\end{equation}
(because $\rmd a=\rmd\rmd\vartheta=0$) and this is in contradiction with
the above assertion that $w\neq0$.~\endproof

The underlying physics for the above no go theorem is the following. In the
continuum, an element of the gauge transformation~$\Lambda(x)$ must be a smooth
function of the coordinate~$x$. In lattice gauge theory, on the other hand,
$\Lambda(x)$ can take an arbitrary value in~$G$ at each site~$x$ and the field
$\Lambda(x)$ can be \emph{arbitrarily random}. Roughly speaking, the space of
lattice gauge fields is much larger than the space of continuum gauge fields,
by the amount of random gauge degrees of freedom. In fact,
$\pi_1(\mathfrak{G})=\pi_{2n+1}(G)\neq0$ in the continuum and the two-parameter
family~(\ref{axsix}) has no continuum analogue (i.e., the configurations are
very random at the cutoff scale). Any \emph{gauge invariant\/} restriction,
like the admissibility~(\ref{twoxseven}), cannot remove such very random
configurations due to gauge degrees of freedom. The gauge degrees of freedom,
even if they are very random, are harmless in usual gauge invariant lattice
theories such as lattice QCD. In our present case of a Weyl fermion in an
anomalous representation, however, gauge symmetry is broken and those random
gauge degrees of freedom cause trouble which we do not encounter in the
continuum. 

The theorem indicates that we have only two physically sensible options.
(It appears that the requirement~4 cannot be sacrificed because it ensures
validity of the Schwinger-Dyson equations.) First is to abandon the
requirement~3 and we allow the real part of the effective action to be gauge
\emph{variant}. This would indeed be the case if one uses the Wilson-Dirac
operator. However, with the Wilson-Dirac operator, clear separation of left and
right chiralities is impossible and we would go back to old controversy on Weyl
nature of the lattice fermion. A manifest gauge invariance of the modulus of
the lattice Weyl determinant is one of main achievements in the recent
developments on lattice chiral gauge theories.

The second option is to abandon the requirement~2 and place some restriction on
the link variables that suppresses random gauge degrees of freedom. This
corresponds to the choice made in this paper; the smooth
condition~(\ref{twoxeleven}) is not gauge invariant and thus evades the theorem.
One may then ask whether the condition~(\ref{twoxeleven}) is enough for a
physically sensible formulation. In the main text, we showed that this is
indeed the case by explicitly constructing a smooth and local fermion
integration measure of a Weyl fermion in the vacuum sector.

\listoftables           
\listoffigures          

\end{document}